Astronomy
&
Astrophysics

# Rotation state, colors, and albedo of the mission-accessible tiny near-Earth asteroid 2001 QJ$_{142}$

Jin Beniyama[1,2,3,*] , Alexey V. Sergeyev[1,4] , David J. Tholen[5], and Marco Micheli[6]

[1] Université Côte d'Azur, Observatoire de la Côte d'Azur, CNRS, Laboratoire Lagrange, Bd de l'Observatoire, CS 34229, 06304 Nice Cedex 4, France
[2] Institute of Astronomy, Graduate School of Science, The University of Tokyo, 2-21-1 Osawa, Mitaka, Tokyo 181-0015, Japan
[3] Department of Astronomy, Graduate School of Science, The University of Tokyo, 7-3-1 Hongo, Bunkyo-ku, Tokyo 113-0033, Japan
[4] Institute of Astronomy, V.N. Karazin Kharkiv National University, 35 Sumska Str., Kharkiv 61022, Ukraine
[5] Institute for Astronomy, University of Hawaii, 2680 Woodlawn Drive, Honolulu, HI 96822, USA
[6] ESA NEO Coordination Centre, Planetary Defence Office, Largo Galileo Galilei 1, 00044 Frascati (RM), Italy

## ABSTRACT

*Context.* Characterizing mission-accessible asteroids using telescopic observations is fundamental for target-selection and planning for spacecraft missions. Near-Earth asteroids on Earth-like orbits are of particular importance for applications such as asteroid mining.
*Aims.* 2001 QJ$_{142}$ is a tiny ($D \leq 100$ m) near-Earth asteroid on an Earth-like orbit with a semimajor axis of 1.06 au, orbital eccentricity of 0.09, and orbital inclination of 3.10°. We aim to characterize 2001 QJ$_{142}$ using ground-based observations with future spacecraft missions in mind.
*Methods.* We performed visible multicolor photometry of 2001 QJ$_{142}$ using the TriCCS on the Seimei 3.8 m telescope in February 2024. We also revisited the images taken with the Suprime-Cam on the Subaru 8.2 m telescope in August 2012.
*Results.* Visible color indices of 2001 QJ$_{142}$ indicate that 2001 QJ$_{142}$ is a C- or X-complex asteroid. We detect a possible fast rotation with a period of about 10 min, which is consistent with a previous report. The geometric albedo of 2001 QJ$_{142}$ is derived to be about 0.3 from a slope of its photometric phase curve, which is consistent with an albedo derived from thermal observations with updated physical quantities. A straightforward interpretation is that 2001 QJ$_{142}$ is either an E- or M-type asteroid, although surface properties of such tiny fast-rotating asteroids are not well understood.
*Conclusions.* We infer that 2001 QJ$_{142}$ is a fast-rotating mission-accessible E- or M-type near-Earth asteroid. More characterizations of tiny asteroids are particularly important for a deeper understanding of their nature.

## 1. Introduction

Characterizing mission-accessible near-Earth asteroids (NEAs) using telescopic observations is fundamental in the target selection of spacecraft missions such as Hayabusa (Fujiwara et al. 2006), Hayabusa2 (Watanabe et al. 2017), and OSIRIS-REx (Lauretta et al. 2017), and this effort has been ongoing for decades (e.g., Mueller et al. 2011; Kuroda et al. 2014; Hasegawa et al. 2018; Thirouin et al. 2016, 2018; Devogèle et al. 2019). By visiting mission-accessible NEAs, we could access primitive surface material from the surface on a short timescale of several years. Moreover, mission-accessible NEAs have attracted attention in the context of the utilization of asteroids as a resource (e.g., Jedicke et al. 2018). The observation conditions and results from telescopic and in situ observations are qualitatively different, and the comparison of physical quantities derived from two independent methods are essential to maximize the scientific outcomes of spacecraft missions.

In recent years, lots of tiny asteroids (with diameters smaller than 100 m) have been discovered by large survey projects, such as the Catalina Sky Survey (CSS; Fuls et al. 2023), the Panoramic Survey Telescope and Rapid Response System

(Pan-STARRS; Chambers et al. 2016), and the Asteroid Terrestrial-impact Last Alert System (ATLAS; Tonry et al. 2018). The smaller the diameter, the larger the number of known asteroids. Thus, more Small asteroids can be explored with less fuel by spacecraft missions. However, only a small fraction of tiny NEAs are characterized, mainly during close approaches. For instance, the fractions of NEAs whose spectral types are estimated are ~1/3, ~1/10, and ~1/100 for 1 km-class, 0.3–1 km-class, and less than 300 m-class bodies, respectively (Perna et al. 2018). This is due to difficulties encountered when observing tiny NEAs, such as large positional uncertainties due to short observational arcs, limited observational windows, and fast motion on the sky.

The NEA 2001 QJ$_{142}$ was discovered by the Lincoln Near-Earth Asteroid Research (LINEAR) project in New Mexico, US, on August 23, 2001[1] (Stokes et al. 2000). 2001 QJ$_{142}$ has a mission-accessible orbit around Earth with a semimajor axis of 1.06 au, an orbital eccentricity of 0.09, and an orbital inclination of 3.10° (e.g., Binzel et al. 2004; Hasnain et al. 2012; Tholen et al. 2012). The impulse required per unit of spacecraft mass

---

* Corresponding author; jbeniyama@oca.eu

[1] https://www.minorplanetcenter.net/mpec/K01/K01Q54.html

**Table 1.** Summary of the observations.

| Obs. date (UTC) | Tel. | Filter | $t_{exp}$ (s) | $N_{img}$ | V (mag) | $\alpha$ (deg) | $\Delta$ (au) | $r_h$ (au) | Air mass |
|---|---|---|---|---|---|---|---|---|---|
| 2012 Aug. 16 13:20:34–13:46:44 | Subaru | $W$-$J$-$VR$ | 90 | 13 | 24.4 | 72.3 | 0.357 | 1.062 | 1.21–1.30 |
| 2024 Feb. 13 12:59:10–13:49:24 | Seimei | $g, r, i$ | 600 | 2 | 20.1 | 20.9 | 0.097 | 1.077 | 1.42–1.57 |
| 2024 Feb. 13 14:06:26–14:56:40 | Seimei | $g, r, z$ | 600 | 6 | 20.1 | 20.9 | 0.097 | 1.077 | 1.37–1.40 |

**Notes.** The observation time in UT at the mid-time of exposure (Obs. date), telescope (Tel.), filters (Filters), total exposure time per frame ($t_{exp}$), and number of images ($N_{img}$) are listed. The predicted $V$-band apparent magnitude ($V$), phase angle ($\alpha$), distance between 2001 QJ$_{142}$ and observer ($\Delta$), and distance between 2001 QJ$_{142}$ and the Sun ($r_h$) at the observation starting time were taken from NASA JPL Horizons on June 16, 2024. Elevations of 2001 QJ$_{142}$ used to calculate the air mass range (Air Mass) are also from NASA JPL Horizons.

to change its status, $\Delta v$, is an important parameter in planning a spacecraft mission. According to the Near-Earth Object Human Space Flight Accessible Targets Study (NHATS; Abell et al. 2012)[2], the estimated $\Delta v$ of 2001 QJ$_{142}$ is about 6 km s$^{-1}$ in the launch window between 2030 and 2035.

Tholen et al. (2012) attempted to obtain recovery observations of 2001 QJ$_{142}$ in 2012 using the Megaprime instrument on the Canada-France-Hawaii Telescope (CFHT) on February 16 and the Tektronix 2048 CCD camera on the University of Hawaii 2.24 m telescope on February 27. Initial examination of the CFHT images did not reveal 2001 QJ$_{142}$, but it was eventually found at about 34 arcmin east of the nominal predicted position. After the recovery, they reanalyzed the CFHT images, and 2001 QJ$_{142}$ was clearly detected in just one of the three 140 s exposures. When blinking the three exposures, one normally expects to see three detections with linear motion. Instead they have one decent detection and two clumps of slightly higher-than-sky pixel values. From large variations in the signal-to-noise ratios in multiple images, they concluded that 2001 QJ$_{142}$ may have a large light curve amplitude and a short rotation period.

Thermal infrared observations of 2001 QJ$_{142}$ were conducted in June 2015 using the Infrared Array Camera (IRAC) on the *Spitzer* Space Telescope (SST; Werner et al. 2004) through the Warm *Spitzer* Cycle 11 Exploration Science program titled NEOSurvey (Trilling et al. 2016). The geometric albedo of 2001 QJ$_{142}$ was estimated to be $p_V = 0.640^{+0.287}_{-0.279}$ ($1\sigma$ uncertainties) using the near-Earth asteroid thermal model (NEATM; Harris 1998). The derived albedo is higher than the average albedo of any spectral type (e.g., Usui et al. 2013). As investigated in Gustafsson et al. (2019), high-albedo values derived from SST measurements are possibly artifacts due to limited light curve coverage.

If 2001 QJ$_{142}$ is truly in the fast-rotating state reported in Tholen et al. (2012), landing a spacecraft on 2001 QJ$_{142}$ will be hard, especially at the equator. Additionally, no characterizations using ground-based telescopes have been reported since 2012, simply due to its faintness and unfavorable observing circumstances. It is worth observing 2001 QJ$_{142}$ again to characterize it as a possible future mission target.

This paper is organized as follows. In Sect. 2, we describe our observations and data reduction procedures. In Sect. 3, the results of multicolor photometry and light curve observations are presented. We discuss the physical properties of 2001 QJ$_{142}$ using all available information in Sect. 4.

## 2. Observations and data reduction

We revisited broadband photometry derived from observations of 2001 QJ$_{142}$ made in Hawaii in 2012 and obtained multicolor photometry of 2001 QJ$_{142}$ from observations made in Japan in 2024. The observing conditions are summarized in Table 1. The predicted $V$-band magnitudes, phase angles, distances between 2001 QJ$_{142}$ and observer, and distances between 2001 QJ$_{142}$ and the Sun in Table 1 were obtained from NASA Jet Propulsion Laboratory (JPL) Horizons[3] using the Python package astroquery (Ginsburg et al. 2019).

### 2.1. Subaru telescope

We searched for images of 2001 QJ$_{142}$ in archive data through the Canadian Astronomy Data Centre (CADC) and found unpublished images of 2001 QJ$_{142}$ taken in 2012. We revisited images taken in two observing runs in Hawaii on August 13 and 16, 2012. These two observations were performed using the Subaru Prime Focus Camera (Suprime-Cam; Miyazaki et al. 2002) on the 8.2 m Subaru telescope (Iye et al. 2004). The field of view is $34' \times 27'$ with a pixel scale of 0.202 arcsec pixel$^{-1}$. In total, 10 exposures with exposure times of 60 s were obtained on August 13, and 17 exposures with exposure times of 90 s on August 16. All images were taken with a wide-band $W$-$J$-$VR$ filter encompassing both the $V$ and $R$ bands[4], and non-sidereal tracking was performed during the observations. The full widths at half maximum (FWHMs) of the point spread functions (PSFs) of the stars in the images from August 13 and 16 registered in the fits header with the keyword SEEING are 0.43 and 0.58 arcsec, respectively. The humidity outside the dome on August 13 and 16 registered in the fits header with the keyword OUT-HUM is ~70.0% and ~10.0%, respectively. Thanks to the recovery by Tholen et al. (2012) and new astrometry measurements from 2024, the locations of 2001 QJ$_{142}$ are very precise. Based on the latest prediction by JPL Horizons as of May 23, 2024, the $3\sigma$ orbital uncertainties of 2001 QJ$_{142}$ on August 13 and 16, 2012, are smaller than 0.2 arcsec, which is comparable to the pixel scale of Subaru/Suprime-Cam. We downloaded all Subaru images from the Subaru-Mitaka-Okayama-Kiso Archive System (SMOKA; Baba et al. 2002) and analyzed the images.

We performed standard bias subtraction, dark subtraction, and flat-fielding with SDFRED2, a useful package for dealing with Subaru/Suprime-Cam images (Ouchi et al. 2004). A master flat image was created using 11 twilight flat images taken on August 14. After the reduction, we searched for 2001 QJ$_{142}$ in the reduced images using the JPL Horizons predictions. We found 2001 QJ$_{142}$ in images taken on August 16 by visual inspection but could not find 2001 QJ$_{142}$ around predicted positions in images taken on August 13. After stacking all ten images taken on August 13, we found 2001 QJ$_{142}$ close to the JPL prediction with a signal-to-noise ratio of ∼5 with an aperture of 4 pix. Thus, we did not use images from August 13 to investigate 2001 QJ$_{142}$'s rotation. The detection of 2001 QJ$_{142}$ with a low signal-to-noise ratio in the August 13 images could be explained by the short exposure times.

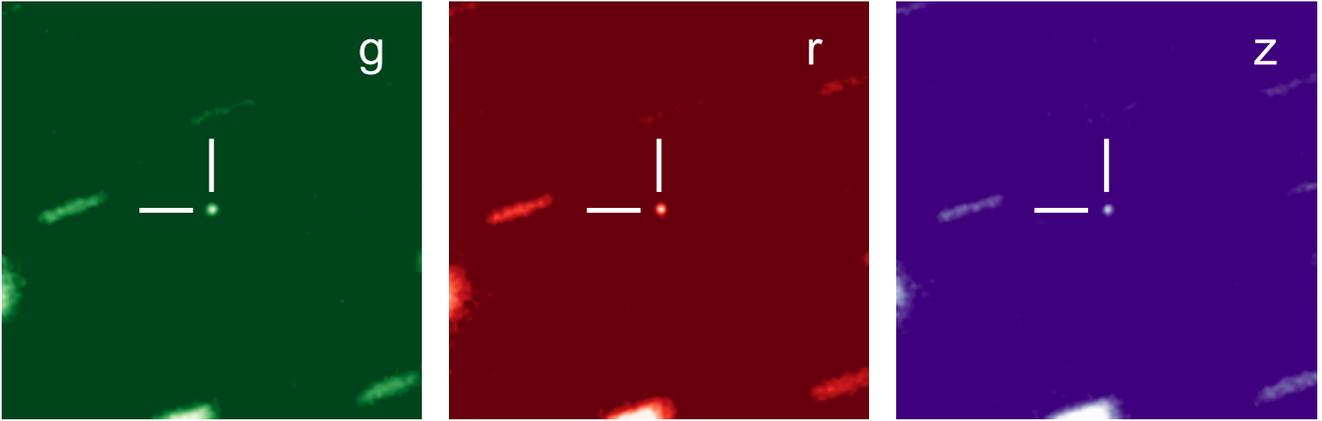

**Fig. 1.** Non-sidereally stacked images of 2001 QJ$_{142}$ in $g$, $r$, and $z$ bands with a total integration time of 600 s from February 13, 2024. Horizontal and vertical bars indicate 2001 QJ$_{142}$. The field of view covers $1.75' \times 1.75'$. North is to the top and east to the left.

We performed circular aperture photometry for 2001 QJ$_{142}$ and two nearby stars using the SExtractor-based `Python` package `sep` (Bertin & Arnouts 1996; Barbary et al. 2017). The 13 images in which 2001 QJ$_{142}$ is far from nearby sources were used for the photometry; the other 4 images were not used. The images of all sources except 2001 QJ$_{142}$ are elongated as a result of the non-sidereal tracking. The aperture radii were set to 4 pix, and annuli from 5 to 7 pix from the center were used to estimate the background level and noise for 2001 QJ$_{142}$; for elongated nearby stars, the aperture radii were set to 12 pix and annuli from 14 to 17 pix from the center were used (see Appendix A). We estimated the final fluxes of 2001 QJ$_{142}$ and nearby star 1 by averaging them and dividing them by those of nearby star 2 (see Appendices A and B for all photometric measurements).

### 2.2. Seimei telescope

We observed 2001 QJ$_{142}$ using the TriColor CMOS Camera and Spectrograph (TriCCS) on the 3.8 m Seimei telescope (Kurita et al. 2020) on February 13, 2024. We simultaneously obtained three-band images in the Pan-STARRS ($g$, $r$, $i$) and ($g$, $r$, $z$) filters (Chambers et al. 2016). The field of view is $12.6' \times 7.5'$ with a pixel scale of 0.350 arcsec pixel$^{-1}$.

Non-sidereal tracking was performed during these observations. The exposure times were set to 10 s. The signal-to-noise ratios in the images are too low to detect 2001 QJ$_{142}$ in a single exposure. We took multiple images with short exposures rather than a single image with a long exposure to avoid having elongated photometric reference stars and also to eliminate the cosmic rays. We performed standard image reduction, including bias subtraction, dark subtraction, and flat-fielding. The astrometry of reference sources from *Gaia* Data Release 2 was performed using the `astrometry.net` software (Lang et al. 2010).

We performed stacking of images before photometry to avoid elongations of the images of 2001 QJ$_{142}$, as shown in the upper panels of Fig. 1 (hereinafter referred to as the non-sidereally stacked image). We stacked 60 successive images with exposure times of 10 s. Since a typical readout time of the CMOS sensors on TriCCS is 0.4 milliseconds, the total integration time is about 600 s, which corresponds to a possible rotation period of 2001 QJ$_{142}$ (see Sect. 3.1). We also stacked images using the World Coordinate System of images corrected using the surrounding sources to suppress the elongations of the images of reference stars, as shown in the lower panels of Fig. 1 (hereinafter referred to as the sidereally stacked image).

We derived colors and magnitudes of 2001 QJ$_{142}$ following the procedure used in Beniyama et al. (2023c,b,a). Cosmic rays were removed with the `Python` package `astroscrappy` (McCully et al. 2018) using the Pieter van Dokkum's `L.A.Cosmic` algorithm (van Dokkum 2001). Circular aperture photometry was performed for 2001 QJ$_{142}$ and the reference stars using the SExtractor-based `Python` package `sep`. The aperture radii were set to 10 pix, which is about 1.7 times larger than the FWHMs of the PSFs of the reference stars on the sidereally stacked images. The photometric results of 2001 QJ$_{142}$ and reference stars were obtained from the non-sidereal and sidereal stacked images, respectively.

## 3. Results

### 3.1. Light curves and rotation period

We obtained the light curves of 2001 QJ$_{142}$ and a nearby star 1 from 13 photometric measurements, as shown in Fig. 2. We performed the periodic analysis using the Lomb–Scargle technique (Lomb 1976; Scargle 1982; VanderPlas 2018). The Lomb–Scargle periodograms with a period range between 240 and 1600 s are shown in Fig. 3, where a peak frequency, $f_{peak}$, is indicated. The minimum and maximum of the period range correspond to twice the sampling rate, which is the sum of the exposure time and overhead time (∼120 s) and the observation arc, respectively. The highest peak around $4.1 \times 10^{-3}$ Hz is an alias corresponding to twice the sampling rate. We show the 90.0, 99.0, and 99.9% confidence levels in the periodogram. The $P$% confidence level represents the probability that a dataset with no signal would lead to a peak of a similar magnitude by $(100 - P)$% assuming the data consist of pure Gaussian noise (VanderPlas 2018).

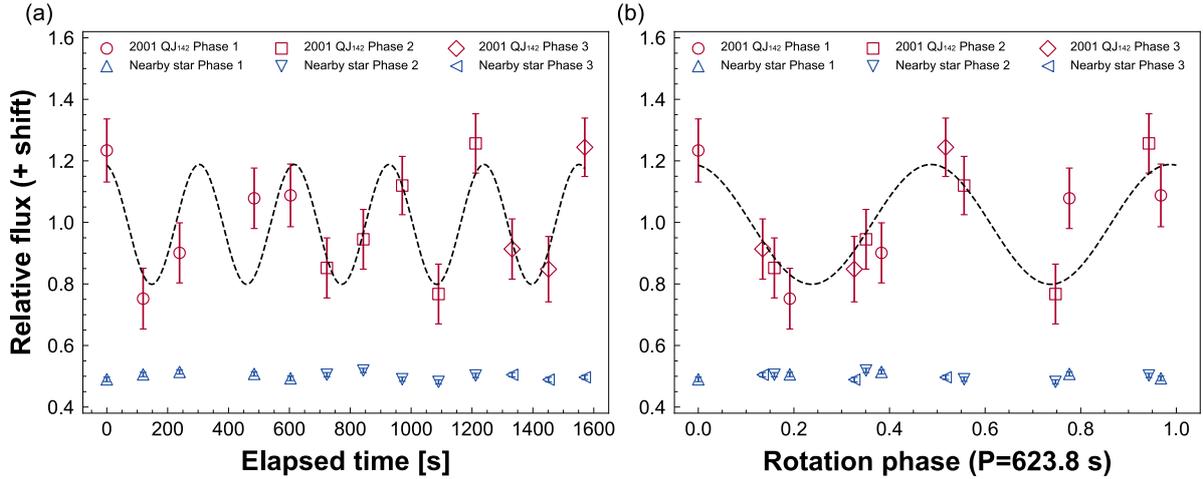

**Fig. 2.** Light curves of 2001 QJ$_{142}$ and nearby star 1. (a) Time-series light curves. (b) Phased light curves. Light curves are relative to nearby star 2 (see Appendix A). Bars indicate the $1\sigma$ uncertainties. A model curve fitted to the light curve with a period of 623.8 s is shown with a dashed line. Relative fluxes of nearby star 1 are shifted by –0.5 for the sake of clarity.

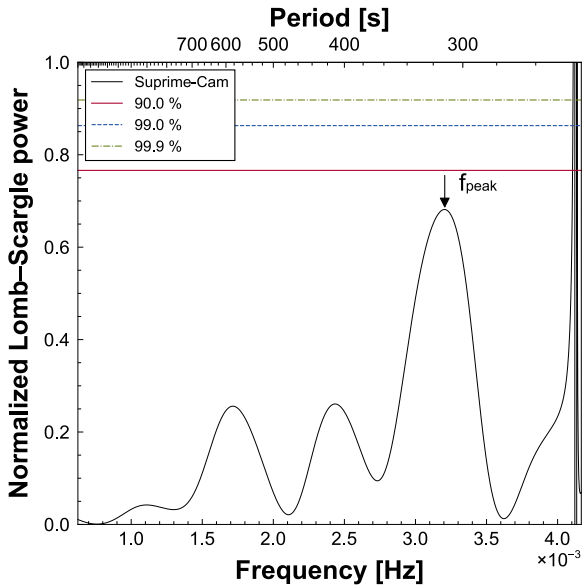

**Fig. 3.** Lomb–Scargle periodogram of 2001 QJ$_{142}$. The number of the harmonics of the model curves is set to unity. A peak frequency corresponding to half of a possible rotation period, 311.9 s, is indicated. Solid, dashed, and dot-dashed horizontal lines show 90.0, 99.0, and 99.9% confidence levels, respectively.

In panels a and b of Fig. 2, we show the model curves with periods of 623.8 s, which corresponds to $f_{peak}/2$. We set the number of harmonics of the model curves to unity. The observed light curves and model curves overlap well, which indicates that 2001 QJ$_{142}$'s rotation period is about 623.8 s. The light curve amplitude of the model curve is 0.43 mag. We cannot rule out the possibility that 2001 QJ$_{142}$ is a non-principal axis rotator (Paolicchi et al. 2002; Pravec et al. 2005); additional observations are necessary to determine this and for a robust estimation of its rotation period.

### 3.2. Colors

We estimated the systematic uncertainties of $g - r$, $r - i$, and $r - z$ colors in our multicolor photometry, $\delta_{g-r}$, $\delta_{r-i}$, and $\delta_{r-z}$, on the basis of the photometric measurements of reference stars as $\delta_{g-r} = 0.02$, $\delta_{r-i} = 0.02$, and $\delta_{r-z} = 0.02$ (Beniyama et al. 2023c). The weighted average colors of 2001 QJ$_{142}$ were derived as $g - r = 0.386 \pm 0.041$, $r - i = 0.148 \pm 0.055$, and $r - z = 0.255 \pm 0.051$. We converted the derived colors to those in the Johnson system for reference using the following equations (Tonry et al. 2012):

$$V - R = 0.605(g - r) + 0.144, \tag{1}$$

$$R - I = 0.018(g - r) + (r - i) + 0.229. \tag{2}$$

We computed the propagated uncertainties of the colors in the Johnson system using the photometric errors and uncertainties in the conversions. The colors correspond to $V - R = 0.377 \pm 0.031$ and $R - I = 0.384 \pm 0.060$.

We present the derived colors of 2001 QJ$_{142}$ in Fig. 4 along with those of asteroids from the latest catalog of moving objects in the archive of the Sloan Digital Sky Survey (SDSS; Sergeyev & Carry 2021). In the catalog, the probabilities of belonging to a certain asteroid complex are assigned for each asteroid. We extracted asteroids with 80% or higher probabilities of belonging to a certain class (except U, which indicates an unknown class) and calculated $g' - r'$, $r' - i'$, and $r' - z'$ in the SDSS system. Then we converted these colors to those in the Pan-STARRS system using the equations from Tonry et al. (2012). Via visual inspection, we see that 2001 QJ$_{142}$ overlaps with C- and X-complex asteroids in the color–color diagrams.

### 3.3. Phase curve

We observed 2001 QJ$_{142}$ at phase angles of about 20.9° with Seimei/TriCCS. We converted the $g$- and $r$-band magnitudes in the Pan-STARRS system to the $V$-band magnitude in the Johnson system using the equations in Tonry et al. (2012) and the $g - r$ and $r - i$ color indices we derived in Sect. 3.2. The phase curve of 2001 QJ$_{142}$ is shown in Fig. 5.

We combined our observational results with those reported in the Minor Planet Center (MPC) Database[5]. There were 119 observations of 2001 QJ$_{142}$ as of July 3, 2024, and 85 of them are reported with observed magnitudes. We used 53 of them calibrated to the $V$-, $R$-, or $G$-band filters; we did not use the other 32, which did not have photometric filters, to minimize the systematic error. We converted the $R$- and $G$-band magnitudes to the $V$-band magnitudes using the equations in Tonry et al. (2012) and Evans et al. (2018), respectively, and the $V - R$ color index we derive in Sect. 3.2.

---

[5] https://www.minorplanetcenter.net/db_search

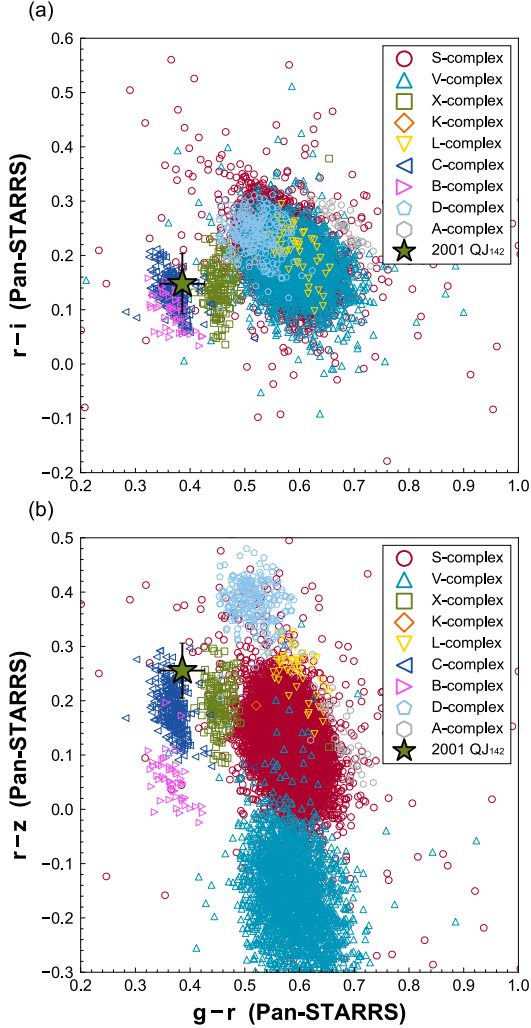

**Fig. 4.** Color–color diagram of (a) $g-r$ vs. $r-i$ and (b) $g-r$ vs. $r-z$. Colors of 2001 QJ$_{142}$ are plotted with star symbols. Bars indicate the $1\sigma$ uncertainties. Asteroids from Sergeyev & Carry (2021) are also plotted: S-complex (circles), V-complex (triangles), X-complex (squares), K-complex (diamonds), L-complex (inverse triangles), C-complex (left-pointing triangles), B-complex (right-pointing triangles), D-complex (pentagons), and A-complex (hexagons).

We derived the absolute magnitude in the $V$ band ($H_V$) and the slope parameter ($G$) using the $H$-$G$ model (Bowell et al. 1989):

$$H_V(\alpha) = H_V - 2.5 \log_{10}\left((1-G)\Phi_1(\alpha) + G\Phi_2(\alpha)\right), \quad (3)$$

where $\alpha$ is a phase angle, and $\Phi_1$ and $\Phi_2$ are phase functions written as follows with a basic function, $W$:

$$\Phi_1(\alpha) = W\left(1 - \frac{0.986 \sin\alpha}{0.119 + 1.341 \sin\alpha - 0.754 \sin^2\alpha}\right) + (1 - W)\exp\left(-3.332 \tan^{0.631}\frac{\alpha}{2}\right), \quad (4)$$

$$\Phi_2(\alpha) = W\left(1 - \frac{0.238 \sin\alpha}{0.119 + 1.341 \sin\alpha - 0.754 \sin^2\alpha}\right) + (1 - W)\exp\left(-1.862 \tan^{1.218}\frac{\alpha}{2}\right), \quad (5)$$

$$W = \exp\left(-90.56 \tan^2\frac{\alpha}{2}\right). \quad (6)$$

We averaged all magnitudes obtained on the same day at the same site. Since uncertainties of magnitudes are not registered in the MPC database, we fit the magnitudes using the $H$-$G$ model and derived best-fit parameters of $H_V = 24.3$ and $G = 0.53$. The model curve with derived parameters is shown in the right panel of Fig. 5. The light curve amplitude of the model curve fitted to the light curve with a period of 623.8 s, 0.43 mag, is also indicated to show the scatter caused by the light curve.

We also used the Shevchenko model (Shevchenko 1996), although we lack low phase angle data (our minimum phase angle is 7.1°):

$$H_V(\alpha) = H_{V,\text{Shev}} + (a\alpha)/(\alpha + 1) + b\alpha, \quad (7)$$

where $H_{V,\text{Shev}}$ is an absolute magnitude in the Shevchenko model, $a$ is the parameter characterizing the opposition effect at small phase angles, and $b$ is the slope of the fitting curve. The $b$ parameter has a tight correlation with the geometric albedo (Belskaya & Shevchenko 2000):

$$b = 0.013 - 0.024 \log_{10} p_V. \quad (8)$$

The derived $b$ of 0.0237 corresponds to a geometric albedo of 0.36. We derived $H_{V,\text{Shev}}$ to be 24.1, which is in good agreement with the $H_V$.

## 4. Discussion

### 4.1. Revised diameter and albedo

The diameter and geometric albedo of 2001 QJ$_{142}$ were derived from the thermal measurement by the SST using the NEATM as $D = 29^{+9}_{-5}$ m ($1\sigma$ uncertainties) and $p_V = 0.640^{+0.287}_{-0.229}$ ($1\sigma$ uncertainties), respectively. The input parameters for the NEATM, $H_V$ and $G$, were assumed to be 23.8 and 0.15, respectively. The measured 4.5 micron flux of 2001 QJ$_{142}$ is $166.2 \pm 12.0$ µJy with a total integration time of 480.0 s (Trilling et al. 2010, 2016)[6]. As already mentioned in the introduction, the derived higher albedo could be an artifact due to the limited light curve coverage of an object with a large light curve amplitude.

We derive the absolute magnitude and slope parameter of 2001 QJ$_{142}$ in Sect. 3.3. We performed the NEATM following the method summarized in Trilling et al. (2016) with the same public code[7] but with updated input parameters ($H_V$ and $G$). This is the only difference between their NEATM and ours. We fit the diameter and albedo of 2001 QJ$_{142}$ with optical and thermal constraints in a Monte Carlo approach, as they did. We performed 10 000 trials with randomized parameters to estimate uncertainties from the distributions. We set the uncertainty of $H_V$ to 0.3 in this analysis. We set a nominal beaming parameter as a function of the phase angle using the following empirical equation (Trilling et al. 2016):

$$\eta(\alpha) = (0.01 \deg^{-1})\alpha + 0.87. \quad (9)$$

---

[6] http://nearearthobjects.nau.edu/spitzerneos.html
[7] https://github.com/mommermi/spitzerneos

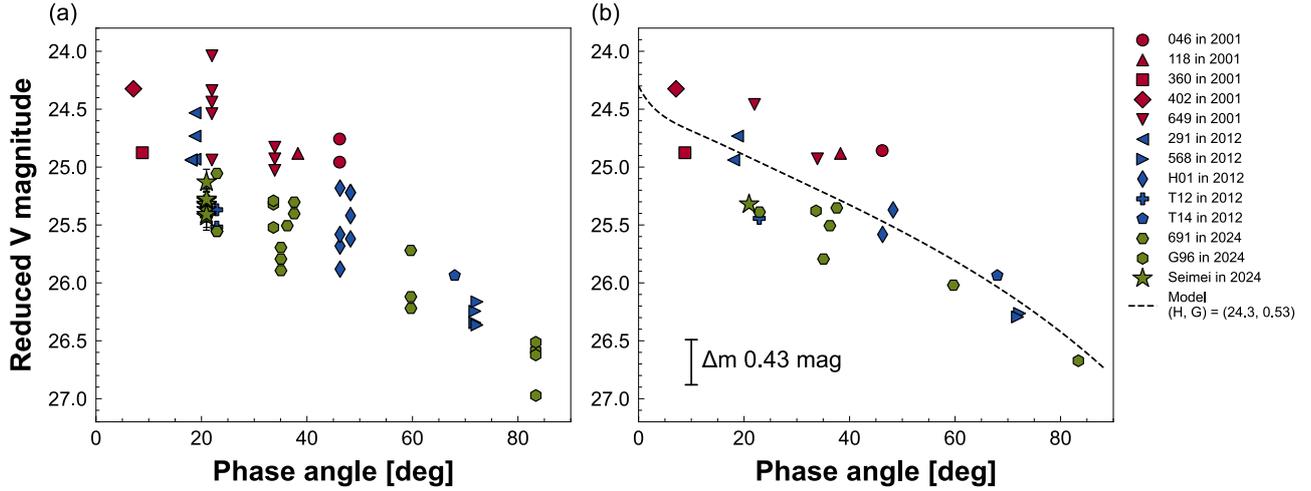

**Fig. 5.** Phase angle dependence of the reduced *V* magnitudes of 2001 QJ$_{142}$. Observations from the MPC database are also plotted with different markers depending on the MPC code after magnitude conversions (see the main text for details). Mean reduced *V*-band magnitudes with Seimei/TriCCS are presented as stars. (a) All magnitudes in the MPC database (except for those without filter information) plotted along with those determined from Seimei/TriCCS data. Bars indicate the 1σ uncertainties. (b) All magnitudes obtained on the same day at the same site, averaged. The fitting model curve with the *H*–*G* model is shown as a dashed line. The light curve amplitude with a period of 623.8 s, 0.43 mag, is indicated.

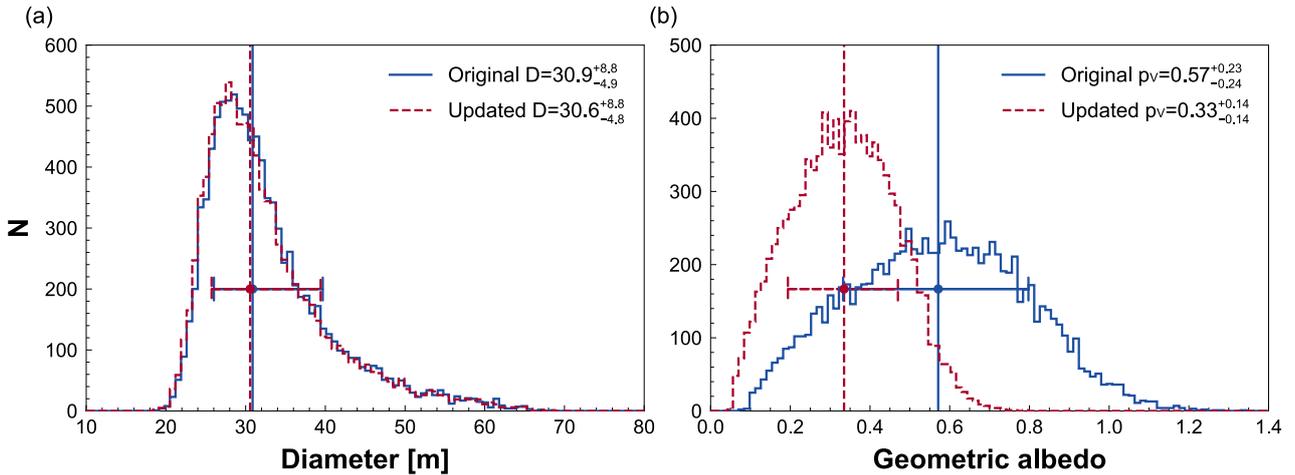

**Fig. 6.** Probability distributions for the (a) diameter and (b) albedo of 2001 QJ$_{142}$. Solutions from the original NEATM and the updated NEATM are shown with solid and dashed lines, respectively. Vertical lines indicate our nominal solutions (the median of the distributions). Horizontal bars indicate the 1σ uncertainties (encompassing 68.3% of the results surrounding the median).

The NEATM results are shown in Fig. 6. The diameter and its 1σ uncertainties were updated from $30.9^{+8.8}_{-4.9}$ m to $30.6^{+8.8}_{-4.8}$ m. The geometric albedo and its 1σ uncertainties were updated from $0.57^{+0.23}_{-0.24}$ to $0.33^{+0.14}_{-0.14}$. We note that the original results are not exactly the same as the published values since the Monte Carlo processes are stochastic. We obtain a geometric albedo for 2001 QJ$_{142}$ of $p_V \sim 0.3$ from both the slope of the photometric phase curve and the thermal modeling. This indicates that our analysis is robust.

### 4.2. Interpretations

We derive a high albedo for 2001 QJ$_{142}$, 0.3. The large slope parameter, *G*, of 0.53 also supports such a high albedo (Shevchenko et al. 2003; MacLennan & Emery 2022). The typical geometric albedos of E and M types in X-complex asteroids are estimated to be $0.559 \pm 0.140$ and $0.175 \pm 0.052$, respectively, whereas the typical geometric albedo of C-complex asteroids is estimated to be ≲0.1 (Usui et al. 2013). Thus, the straightforward interpretation is that 2001 QJ$_{142}$ is a mission-accessible E- or M-type NEA. Considering a few tens of percent uncertainties in the derived albedo, we could not determine whether 2001 QJ$_{142}$ is an E or M type. No in situ exploration of E- and M-type NEAs has ever been performed.

Albedos of some C-complex asteroids have been estimated to be larger than the typical values determined from thermal infrared measurements (Mueller et al. 2011; Thomas et al. 2011). This so-called class-albedo discrepancy indicates that some assumptions in thermal modeling may be inappropriate. As for the kilometer-sized C-complex NEA (3671) Dionysus, Mueller et al. (2011) originally derived its nominal albedo to be 0.55, much higher than the typical albedo for a C-complex asteroid. They concluded that this is due to the inappropriate assumption of the beaming parameter in their thermal modeling, and they revised a nominal albedo from 0.55 to 0.178, which is consistent with a C-complex classification. We also derived the diameter and albedo of 2001 QJ$_{142}$ using the fast rotating model (FRM; e.g., Lebofsky & Spencer 1989) to be $42^{+2}_{-2}$ m and $0.18^{+0.04}_{-0.04}$, respectively. The FRM assumes that a temperature distribution is isothermal in longitude and depends only on latitude, which is different from the temperature distribution in NEATM. It is possible that the FRM is more appropriate for a fast-rotating asteroid such as 2001 $QJ_{142}$. However, the important point is that these models have not yet been tested for such tiny ($D \sim 30$ m) and fast-rotating ($P \sim 10$ min) asteroids. Recent studies show that the thermal inertia of fast-rotating asteroids could be lower than that of large asteroids (Fenucci et al. 2021, 2023), but this is still a matter of debate since the sample is very limited. Thus, Eq. (9) could be inappropriate for 2001 $QJ_{142}$, or in other words, it is still unclear what thermal model is appropriate for such tiny fast-rotating asteroids. Further investigation, while valuable, is beyond the scope of this paper since we lack knowledge of the surface environment of tiny fast-rotating asteroids.

### 4.3. Nongravitational acceleration

We performed an accurate orbit determination for the object using all the existing astrometry, combining all positional records currently available from the MPC with our high-precision astrometric measurements of the datasets obtained by our team in 2012. For our astrometric positions, we derived our own estimate of the astrometric uncertainties, while for the MPC dataset we relied on uncertainties provided in the ADES astrometric format, when available. For cases where no astrometric uncertainties could be obtained, we assumed the default values from the error model by Vereš et al. (2017) and a conservative default of ±0.6 arcsec for stations not included in the model. An overall fit of the entire dataset, extending from August 23, 2001, to April 5, 2024, with a purely gravitational model leads to unreasonably high astrometric residuals over the entire arc, and a reduced chi-square metric in excess of 4, clearly highlighting the existence of additional perturbations. To model such perturbations, we included a tangential acceleration, $A_2$, following the Yarkovsky formalism (Farnocchia et al. 2013; Fenucci et al. 2024), which lowered the reduced chi-square to a value of 0.99 when including all available observations. The best-fit orbital determination corresponds to a transverse acceleration, $A_2 = (-8.47 \pm 0.13) \times 10^{-14}$ au d$^{-2}$. A further attempt to improve the orbital fit with the addition of a radial and an out-of-plane acceleration did not lead to a statistically significant improvement in the quality of the orbital fit, with neither of the two added components being detectable at a significance level above $5\sigma$.

The negative $A_2$ value indicates that the semimajor axis of 2001 $QJ_{142}$ is decreasing, implying a possible retrograde rotation, which could be of importance. However, the thermal properties of tiny fast-rotating asteroids remain poorly understood, making it unclear if the correlation holds for such objects. There are currently 18 asteroids with periods faster than 1 h and a nonzero $A_2$ in the JPL Small-Body DataBase (SBDB)[8]. To the best of our knowledge, the rotation direction has been estimated for 3 out of these 18 asteroids: (54509) YORP (Taylor et al. 2007), 2011 MD (Mommert et al. 2014), and 2012 $TC_4$ (Lee et al. 2021). 2011 MD and 2012 $TC_4$ are more complex objects because they are small enough ($D \lesssim 10$ m) to have a significant solar radiation pressure signature. Additionally, if their shapes are sufficiently irregular, some of the acceleration may include tangential components, altering the overall Yarkovsky effect signal. Thus, (54509) YORP ($D \sim 100$ m) is the only one for which we can safely assume to be generated almost entirely by Yarkovsky effect. (54509) YORP has an obliquity of 173 deg (Taylor et al. 2007) and a negative $A_2$ of $(-8.72 \pm 3.07) \times 10^{-14}$ au d$^{-2}$ (JPL SBDB). Despite the small sample size, the correlation holds for (54509) YORP, which may indicate a retrograde rotation for 2001 $QJ_{142}$.

The next observing windows of 2001 $QJ_{142}$ will take place in October 2024 ($V \sim 20$ mag) and November 2035 ($V \sim 21$ mag). Additional observations with wide wavelength coverage are encouraged to deepen our knowledge of 2001 $QJ_{142}$. In the coming years, a number of new and "lost" asteroids will be discovered and recovered, respectively, by the *Rubin* Observatory Legacy Survey of Space and Time (LSST) in Chile (Ivezić et al. 2019; Ivezić & Ivezić 2021). The LSST is expected to observe more than 5 million asteroids, including 100000 NEAs, during its ten-year survey using the Simonyi Survey Telescope with an effective aperture of 6.4 m. The statistical studies of tiny NEAs via dedicated follow-up observations will deepen our knowledge of tiny bodies.

## 5. Conclusions

We conducted visible multicolor photometry of 2001 $QJ_{142}$ with Seimei/TriCCS in February 2024. We also revisited Subaru/Suprime-Cam images taken in August 2012. We derive visible color indices for 2001 $QJ_{142}$ of $g - r = 0.386 \pm 0.041$, $r - i = 0.148 \pm 0.055$, and $r - z = 0.255 \pm 0.051$, which indicate that 2001 $QJ_{142}$ is a C- or X-complex asteroid. We confirm a fast rotation with a period of about 10 min, and a negative $A_2$ value implies that 2001 $QJ_{142}$ is a retrograde rotator. A geometric albedo for 2001 $QJ_{142}$ of $\sim 0.3$ was derived from a slope of its photometric phase curve; this is consistent with an albedo derived from thermal observations with updated physical quantities. We infer that 2001 $QJ_{142}$ is a mission-accessible E- or M-type NEA, although some of our assumptions may be inappropriate, due to weak gravity and/or fast rotation. More characterizations of tiny asteroids are particularly important for a deeper understanding of their nature.

*Acknowledgements.* The authors are grateful to Dr. Tsuyoshi Terai for helpful comments on the data reduction of Suprime-Cam images. We thank Dr. Shinichi Kinoshita and Mr. Sorato Wada for observing assistance. We are grateful to the referee for valuable comments that significantly improved the quality of this manuscript. The authors thank the TriCCS developer team (which has been supported by the JSPS KAKENHI grant Nos. JP18H05223, JP20H00174, and JP20H04736, and by NAOJ Joint Development Research). This research used the facilities of the Canadian Astronomy Data Centre operated by the National Research Council of Canada with the support of the Canadian Space Agency. Based in part on data collected at Subaru Telescope and obtained from the SMOKA, which is operated by the Astronomy Data Center, National Astronomical Observatory of Japan. The Pan-STARRS1 Surveys (PS1) and the PS1 public science archive have been made possible through contributions by the Institute for Astronomy, the University of Hawaii, the Pan-STARRS Project Office, the Max-Planck Society and its participating institutes, the Max Planck Institute for Astronomy, Heidelberg and the Max Planck Institute for Extraterrestrial Physics, Garching, The Johns Hopkins University, Durham University, the University of Edinburgh, the Queen's University Belfast, the Harvard-Smithsonian Center for Astrophysics, the Las Cumbres Observatory Global Telescope Network Incorporated, the National Central University of Taiwan, the Space Telescope Science Institute, the National Aeronautics and Space Administration under grant No. NNX08AR22G issued through the Planetary Science Division of the NASA Science Mission Directorate, the National Science Foundation grant No. AST-1238877, the University of Maryland, Eotvos Lorand University (ELTE), the Los Alamos National Laboratory, and the Gordon and Betty Moore Foundation. This work was supported by JSPS KAKENHI Grant Numbers JP22K21344 and JP23KJ0640.

# References

Abell, P. A., Barbee, B. W., Mink, R. G., et al. 2012, in 43rd Annual Lunar and Planetary Science Conference, Lunar and Planetary Science Conference, 2842

Baba, H., Yasuda, N., Ichikawa, S.-I., et al. 2002, in Astronomical Society of the Pacific Conference Series, 281, Astronomical Data Analysis Software and Systems XI, eds. D. A. Bohlender, D. Durand, & T. H. Handley, 298

Barbary, K., Boone, K., Craig, M., Deil, C., & Rose, B. 2017, https://doi.org/10.5281/zenodo.896928

Belskaya, I. N., & Shevchenko, V. G. 2000, Icarus, 147, 94

Beniyama, J., Ohsawa, R., Avdellidou, C., et al. 2023a, AJ, 166, 229

Beniyama, J., Sako, S., Ohtsuka, K., et al. 2023b, ApJ, 955, 143

Beniyama, J., Sekiguchi, T., Kuroda, D., et al. 2023c, PASJ, 75, 297

Bertin, E., & Arnouts, S. 1996, A&AS, 117, 393

Binzel, R. P., Perozzi, E., Rivkin, A. S., et al. 2004, Meteor. Planet. Sci., 39, 351

Bowell, E., Hapke, B., Domingue, D., et al. 1989, in Asteroids II, eds. R. P. Binzel, T. Gehrels, & M. S. Matthews (Tucson, AZ: Univ. Arizona Press), 524

Chambers, K. C., Magnier, E. A., Metcalfe, N., et al. 2016, arXiv e-prints [arXiv:1612.05560]

Devogèle, M., Moskovitz, N., Thiroun, A., et al. 2019, AJ, 158, 196

Evans, D. W., Riello, M., De Angeli, F., et al. 2018, A&A, 616, A4

Farnocchia, D., Chesley, S. R., Vokrouhlický, D., et al. 2013, Icarus, 224, 1

Fenucci, M., Novaković, B., Vokrouhlický, D., & Weryk, R. J. 2021, A&A, 647, A61

Fenucci, M., Novaković, B., & Marčeta, D. 2023, A&A, 675, A134

Fenucci, M., Micheli, M., Gianotto, F., et al. 2024, A&A, 682, A29

Fujiwara, A., Kawaguchi, J., Yeomans, D. K., et al. 2006, Science, 312, 1330

Fuls, D., Christensen, E., Fay, D., et al. 2023, in AAS/Division for Planetary Sciences Meeting Abstracts, 55, 405.08

Ginsburg, A., Sipőcz, B. M., Brasseur, C. E., et al. 2019, AJ, 157, 98

Gustafsson, A., Trilling, D. E., Mommert, M., et al. 2019, AJ, 158, 67

Harris, A. W. 1998, Icarus, 131, 291

Hasegawa, S., Kuroda, D., Kitazato, K., et al. 2018, PASJ, 70, 114

Hasnain, Z., Lamb, C. A., & Ross, S. D. 2012, Acta Astronaut., 81, 523

Ivezić, V., & Ivezić, Ž. 2021, Icarus, 357, 114262

Ivezić, Ž., Kahn, S. M., Tyson, J. A., et al. 2019, ApJ, 873, 111

Iye, M., Karoji, H., Ando, H., et al. 2004, PASJ, 56, 381

Jedicke, R., Bolin, B. T., Bottke, W. F., et al. 2018, Front. Astron. Space Sci., 5, 13

Kurita, M., Kino, M., Iwamuro, F., et al. 2020, PASJ, 72, 48

Kuroda, D., Ishiguro, M., Takato, N., et al. 2014, PASJ, 66, 51

Lang, D., Hogg, D. W., Mierle, K., Blanton, M., & Roweis, S. 2010, AJ, 139, 1782

Lauretta, D. S., Balram-Knutson, S. S., Beshore, E., et al. 2017, Space Sci. Rev., 212, 925

Lebofsky, L. A., & Spencer, J. R. 1989, in Asteroids II, eds. R. P. Binzel, T. Gehrels, & M. S. Matthews, 128

Lee, H.-J., Ďurech, J., Vokrouhlický, D., et al. 2021, AJ, 161, 112

Lomb, N. R. 1976, Ap&SS, 39, 447

MacLennan, E. M., & Emery, J. P. 2022, psj, 3, 47

McCully, C., Crawford, S., Kovacs, G., et al. 2018, https://doi.org/10.5281/zenodo.1482019

Miyazaki, S., Komiyama, Y., Sekiguchi, M., et al. 2002, PASJ, 54, 833

Mommert, M., Farnocchia, D., Hora, J. L., et al. 2014, ApJ, 789, L22

Mueller, M., Delbo', M., Hora, J. L., et al. 2011, AJ, 141, 109

Ouchi, M., Shimasaku, K., Okamura, S., et al. 2004, ApJ, 611, 660

Paolicchi, P., Burns, J. A., & Weidenschilling, S. J. 2002, Side Effects of Collisions: Spin Rate Changes, Tumbling Rotation States, and Binary Asteroids, eds. J. Bottke, W. F., A. Cellino, P. Paolicchi, & R. P. Binzel, 517

Perna, D., Barucci, M. A., Fulchignoni, M., et al. 2018, Planet. Space Sci., 157, 82

Pravec, P., Harris, A. W., Scheirich, P., et al. 2005, Icarus, 173, 108

Scargle, J. D. 1982, ApJ, 263, 835

Sergeyev, A. V., & Carry, B. 2021, A&A, 652, A59

Shevchenko, V. G. 1996, in Lunar and Planetary Science Conference, 27, 1193

Shevchenko, V. G., Krugly, Y. N., Chiorny, V. G., Belskaya, I. N., & Gaftonyuk, N. M. 2003, Planet. Space Sci., 51, 525

Stokes, G. H., Evans, J. B., Viggh, H. E. M., Shelly, F. C., & Pearce, E. C. 2000, Icarus, 148, 21

Taylor, P. A., Margot, J.-L., Vokrouhlický, D., et al. 2007, Science, 316, 274

Thirouin, A., Moskovitz, N., Binzel, R. P., et al. 2016, AJ, 152, 163

Thirouin, A., Moskovitz, N. A., Binzel, R. P., et al. 2018, ApJS, 239, 4

Tholen, D. J., Micheli, M., & Elliott, G. T. 2012, in AAS/Division for Planetary Sciences Meeting Abstracts, 44, 111.12

Thomas, C. A., Trilling, D. E., Emery, J. P., et al. 2011, AJ, 142, 85

Tonry, J. L., Stubbs, C. W., Lykke, K. R., et al. 2012, ApJ, 750, 99

Tonry, J. L., Denneau, L., Heinze, A. N., et al. 2018, PASP, 130, 064505

Trilling, D. E., Mueller, M., Hora, J. L., et al. 2010, AJ, 140, 770

Trilling, D. E., Mommert, M., Hora, J. L., et al. 2016, AJ, 152, 172

Usui, F., Kasuga, T., Hasegawa, S., et al. 2013, ApJ, 762, 56

van Dokkum, P. G. 2001, PASP, 113, 1420

VanderPlas, J. T. 2018, ApJS, 236, 16

Vereš, P., Farnocchia, D., Chesley, S. R., & Chamberlin, A. B. 2017, Icarus, 296, 139

Watanabe, S.-i., Tsuda, Y., Yoshikawa, M., et al. 2017, Space Sci. Rev., 208, 3

Werner, M. W., Roellig, T. L., Low, F. J., et al. 2004, ApJS, 154, 1

# Appendix A: Validation of photometric measurements with Subaru/Suprime-Cam

We present the photometric results of 2001 QJ$_{142}$ and two nearby stars in Figs. A.1–A.4 for validation purposes.

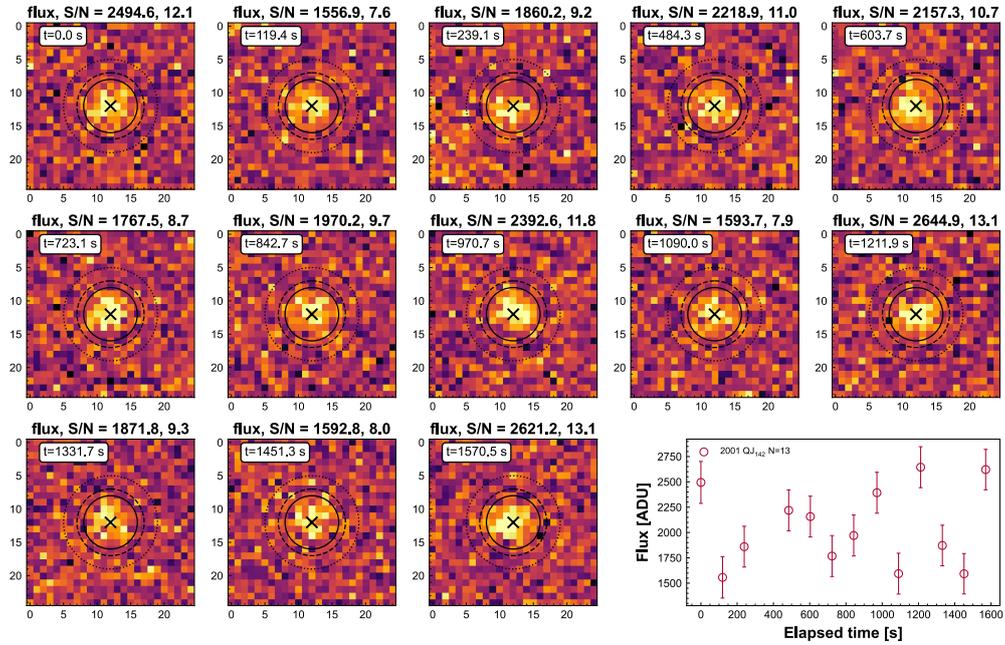

**Fig. A.1.** Cutout images of 2001 QJ$_{142}$ taken with Subaru/Suprime-Cam. Crosses correspond to centers of apertures. Apertures with radii of pixels are shown as solid lines. The inner and outer edges of annuli used to estimate background levels are shown as dashed and dotted lines, respectively. Fluxes estimated using aperture photometry are shown in the lower-right corner.

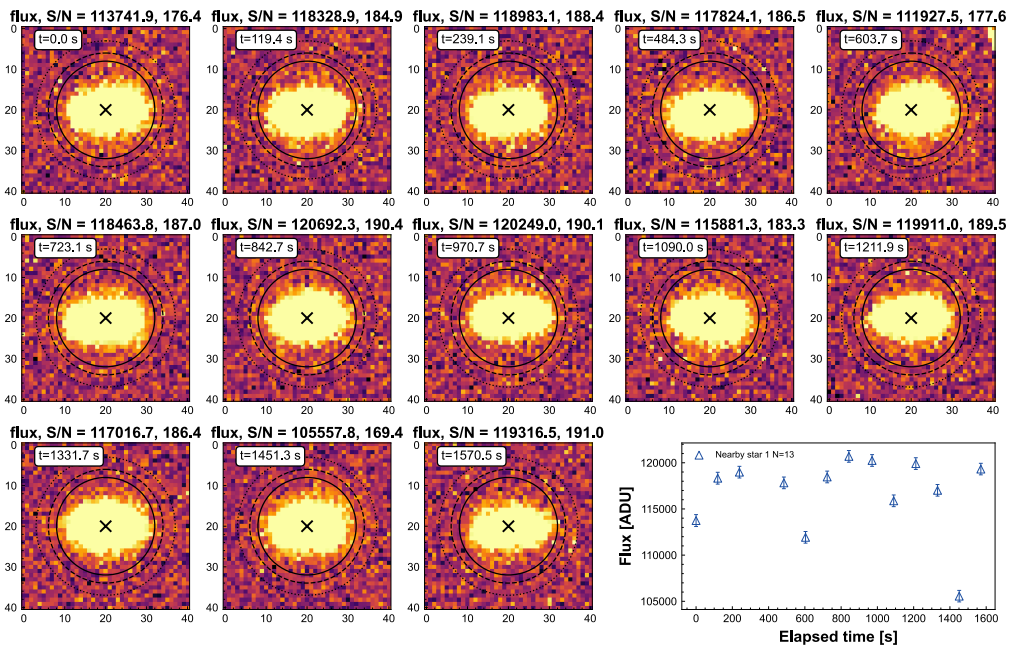

**Fig. A.2.** Same as Fig. A.1 but for nearby star 1.

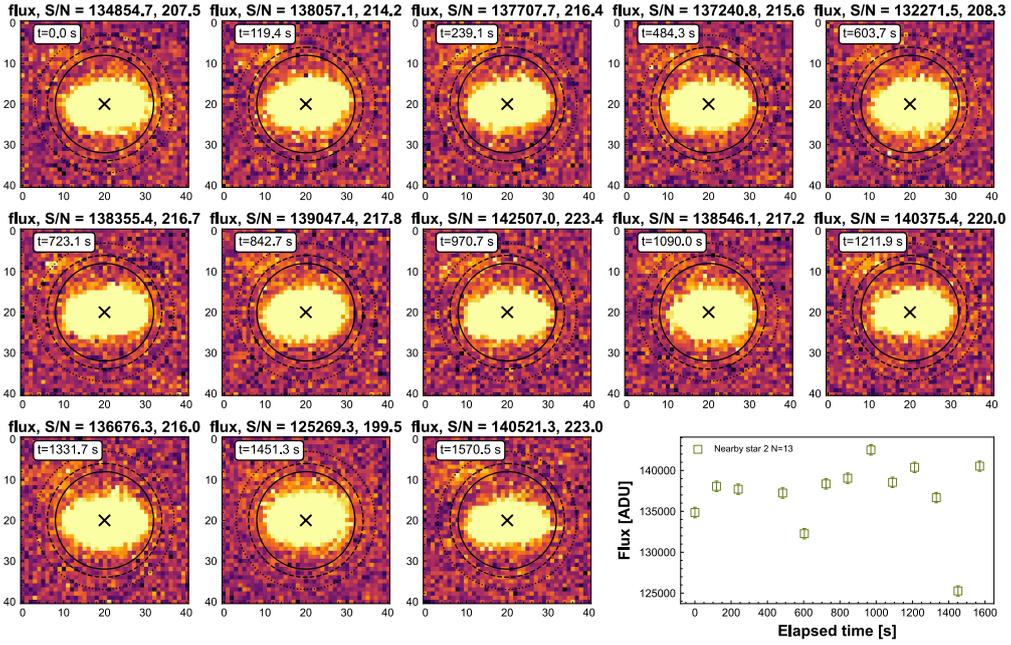

**Fig. A.3.** Same as Fig. A.1 but for nearby star 2.

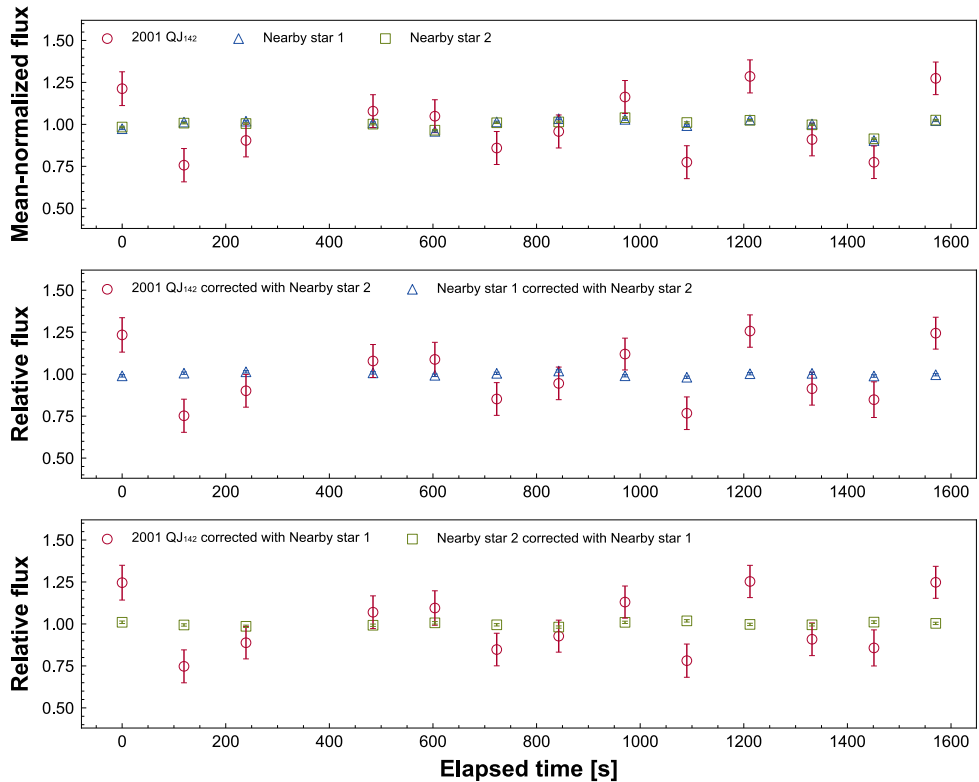

**Fig. A.4.** Light curves of 2001 QJ142 and two nearby stars. Top: Fluxes of 2001 QJ$_{142}$ and two nearby stars. Measured fluxes are normalized by the mean such that the averages are unity. Bars indicate the $1\sigma$ uncertainties. Uncertainties of two nearby stars are hard to see due to scale effects. Middle: Light curves of 2001 QJ$_{142}$ and nearby star 1 corrected using nearby star 2. Relative fluxes of 2001 QJ$_{142}$ and nearby star 1 are presented as circles and triangles, respectively. Bars indicate the $1\sigma$ uncertainties. Bottom: Light curves of 2001 QJ$_{142}$ and nearby star 2 corrected using nearby star 1. Relative fluxes of 2001 QJ$_{142}$ and nearby star 2 are presented as circles and squares, respectively. Bars indicate the $1\sigma$ uncertainties.

## Appendix B: Photometric measurements

We summarize all photometric measurements of 2001 QJ$_{142}$ and two nearby stars in Tables B.1 and B.2.

**Table B.1.** Photometric measurements taken with Subaru/Suprime-Cam.

| JD (day) | Elapsed time (s) | $F_{QJ142}$ | $F_{star1}$ | $F_{star2}$ |
|---|---|---|---|---|
| 2456156.05596 | 0.0 | 2494.6 ± 207.0 | 113741.9 ± 644.8 | 134854.7 ± 649.9 |
| 2456156.05734 | 119.4 | 1556.9 ± 204.1 | 118328.9 ± 639.8 | 138057.1 ± 644.6 |
| 2456156.05872 | 239.1 | 1860.2 ± 201.5 | 118983.1 ± 631.6 | 137707.7 ± 636.2 |
| 2456156.06156 | 484.3 | 2218.9 ± 201.9 | 117824.1 ± 631.7 | 137240.8 ± 636.5 |
| 2456156.06294 | 603.7 | 2157.3 ± 201.8 | 111927.5 ± 630.1 | 132271.5 ± 635.1 |
| 2456156.06433 | 723.1 | 1767.5 ± 202.1 | 118463.8 ± 633.6 | 138355.4 ± 638.5 |
| 2456156.06571 | 842.7 | 1970.2 ± 202.2 | 120692.3 ± 633.9 | 139047.4 ± 638.4 |
| 2456156.06719 | 970.7 | 2392.6 ± 202.1 | 120249.0 ± 632.5 | 142507.0 ± 638.0 |
| 2456156.06857 | 1090.0 | 1593.7 ± 201.8 | 115881.3 ± 632.3 | 138546.1 ± 637.8 |
| 2456156.06998 | 1211.9 | 2644.9 ± 202.4 | 119911.0 ± 632.9 | 140375.4 ± 637.9 |
| 2456156.07137 | 1331.7 | 1871.8 ± 200.4 | 117016.7 ± 627.9 | 136676.3 ± 632.8 |
| 2456156.07276 | 1451.3 | 1592.8 ± 199.5 | 105557.8 ± 623.1 | 125269.3 ± 628.0 |
| 2456156.07413 | 1570.5 | 2621.2 ± 199.7 | 119316.5 ± 624.8 | 140521.3 ± 630.1 |

**Notes.** The observation time in Julian day (JD), the elapsed time (Elapsed time), the flux and its uncertainty of 2001 QJ$_{142}$ ($F_{QJ142}$), the flux and its uncertainty of nearby star 1 ($F_{star1}$), and the flux and its uncertainty of nearby star 2 ($F_{star2}$) are listed. Julian day is the mid-time of exposure, and light time correction is not applied.

**Table B.2.** Photometric measurements of 2001 QJ$_{142}$ taken with Seimei/TriCCS.

| JD (day) | $\alpha$ (deg) | $g\text{-}r$ (mag) | $r\text{-}i$ (mag) | $r\text{-}z$ (mag) | $V$ (mag) |
|---|---|---|---|---|---|
| 2460354.04110 | 20.9 | 0.444 ± 0.105 | 0.034 ± 0.076 | – | 25.130 ± 0.112 |
| 2460354.07598 | 20.9 | 0.472 ± 0.108 | 0.245 ± 0.070 | – | 25.332 ± 0.115 |
| 2460354.08781 | 20.9 | 0.362 ± 0.104 | – | 0.098 ± 0.129 | 25.322 ± 0.110 |
| 2460354.09479 | 20.9 | 0.230 ± 0.107 | – | 0.450 ± 0.116 | 25.428 ± 0.118 |
| 2460354.10176 | 20.9 | 0.425 ± 0.101 | – | 0.114 ± 0.124 | 25.369 ± 0.106 |
| 2460354.10875 | 20.9 | 0.543 ± 0.095 | – | 0.189 ± 0.104 | 25.291 ± 0.096 |
| 2460354.11572 | 20.9 | 0.379 ± 0.091 | – | 0.298 ± 0.102 | 25.278 ± 0.096 |
| 2460354.12270 | 20.9 | 0.208 ± 0.101 | – | 0.340 ± 0.116 | 25.410 ± 0.111 |

**Notes.** The observation time in Julian day (JD), phase angle ($\alpha$), $g\text{-}r$ color, $r\text{-}i$ color, and $r\text{-}z$ color, and the $V$-band magnitude are listed. Julian day is the mid-time of exposure, and light time correction is applied.